# Observation of anomalous Floquet non-Abelian topological insulators

Huahui Qiu,[†] Shuaishuai Tong,[†] Qicheng Zhang, Kun Zhang, and Chunyin Qiu[*]

Key Laboratory of Artificial Micro- and Nano-Structures of Ministry of Education and
School of Physics and Technology, Wuhan University, Wuhan 430072, China

[†]These authors contributed equally: Huahui Qiu, Shuaishuai Tong

[*]To whom correspondence should be addressed: cyqiu@whu.edu.cn

*Abstract.* Non-Abelian topological phases, which go beyond traditional Abelian topological band theory, are garnering increasing attention[1-4]. This is further spurred by periodic driving[5-10], leading to predictions of many novel multi-gap Floquet topological phases, including anomalous Euler and Dirac string phases induced by non-Abelian Floquet braiding[11], as well as Floquet non-Abelian topological insulators (FNTIs) that exhibit multifold bulk-edge correspondence[12]. Here, we report the first experimental realization of anomalous FNTIs, which demonstrate topological edge modes in all three gaps despite having a trivial bulk charge. Concretely, we construct an experimentally feasible one-dimensional three-band Floquet model and implement it in acoustics by integrating time-periodic coupling circuits to static acoustic crystals. Furthermore, we observe counterintuitive topological interface modes in the domain-wall formed by an anomalous FNTI and its counterpart with swapped driving sequences—modes previously inaccessible in Floquet Abelian systems. Our work paves the way for further experimental exploration of the uncharted non-equilibrium topological physics.

The discovery of topological insulators has revolutionized our understanding of material properties through the lens of symmetry and topology[13-17]. The global invariants underlying these phases are often characterized by Abelian charges like Chern numbers or winding numbers. Recently, the concept of non-Abelian charges has been introduced to $\mathcal{PT}$-symmetric systems with multiple intertwined bandgaps[1-4,18]. For instance, a one-dimensional (1D) three-band topological insulator can be described by the non-Abelian quaternion group $Q_8 = \{+1, \pm i, \pm j, \pm k, -1\}$, which obeys the fundamental multiplication rules $i^2 = j^2 = k^2 = ijk = -1$ and $ij = -ji$, $ik = -ki$, $jk = -kj$. In this context, the charge $q = +1$ signifies a trivial phase without topological edge modes (TEMs) in any bandgap. In contrast to their Abelian counterparts, non-Abelian topological phases display more intricate behaviors, such as trajectory-dependent nodal point collisions in two-dimensional semimetals, admissible nodal line configurations in three-dimensional semimetals, and multi-gap TEMs governed by nontrivial quaternion charges[1-4,18]. Experimentally, non-Abelian phases with tangled multi-gap topology have been realized across a variety of platforms, including photonic systems[19-22], acoustic crystals[23-26], and transmission line networks[27,28].

Floquet systems are out-of-equilibrium quantum states that evolve under periodic driving[5-10,29-32]. As a powerful tool for manipulating band structures, Floquet engineering enables novel topological phenomena without static analogs, such as anomalous chiral edge modes even when all bulk bands carry trivial Chern numbers[10]. Very recently, the synergy of Floquet setting and non-Abelian topology has triggered many unprecedented topological phases[11,12,33]. Notable examples include anomalous Euler (Dirac string) phases[11] and Floquet non-Abelian topological insulators (FNTIs)[12]. Intriguingly, the latter exhibit a multifold bulk-edge correspondence governed by the multiplication rule of the quaternion group $Q_8$ (see Fig. 1). Mathematically, the bulk charge $q$ of a 1D FNTI equals the ordered product of the (quaternion) charges $\overline{q}_m$ of all phase-band singularities (i.e., $q = \prod_m \overline{q}_m$), each resulting in a change in the mass term and the emergence of an in-gap TEM according to Jackiw-Rebbi's argument[12,34]. This enables multifold bulk-edge correspondence, as a bulk charge $q$ can be expressed in different multiplicative forms, each corresponding to a specific TEM configuration. For example, in the absence of phase-band singularities, the trivial charge $q = +1$ does not contribute any TEM. In contrast, the charge $q = +1$ represented by $\overline{j} \cdot \overline{i} \cdot \overline{k}$ gives rise to TEMs in all bandgaps, thereby defining an anomalous FNTI.

To date, although Floquet Abelian systems have been implemented across various platforms[35-44], Floquet non-Abelian topological phenomena have yet to be observed in any experiment. This is primarily due to the significant challenges in realizing real-time dynamic couplings and tracking the complicated drive-induced multi-gap interactions in existing experimental platforms. Here, we report the first experimental implementation of the highly elusive FNTIs using our acoustic platform. By adopting an electric circuit-driven dynamic coupling strategy, we successfully realize the anomalous FNTI ($q = +1$) that hosts TEMs in all three gaps, alongside a comparative FNTI ($q = \pm j$) where the $\pi$-gap TEM vanishes. Unambiguously, we demonstrate these two FNTIs by characterizing their quasienergy spectra and TEM patterns, through both zero-order and high-order harmonic signals. More intriguingly, we observe a unique interface response at the domain wall between the anomalous FNTI and its counterpart with swapped driving sequences—another hallmark manifestation of non-Abelian dynamics in multi-gap Floquet systems.

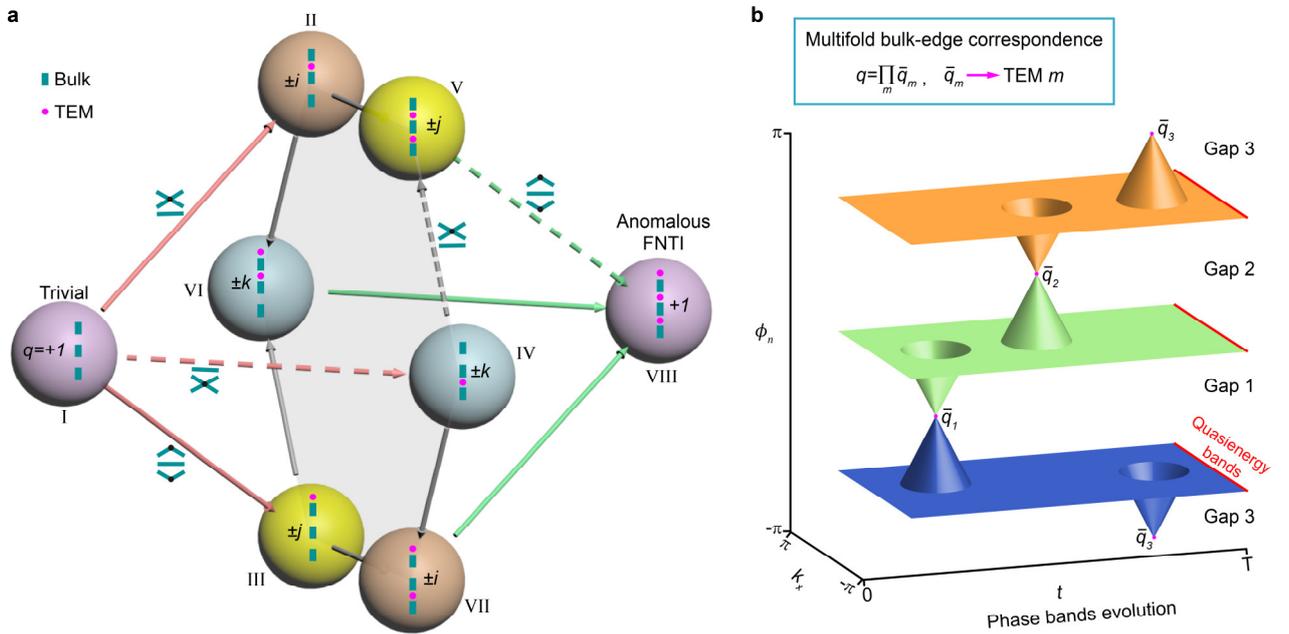

**Fig. 1 | Multifold bulk-edge correspondence of 1D three-band FNTIs. a** Topological transitions and the resulting TEM configurations. Starting from the trivial phase I with $q = +1$, the system undergoes gap closure and reopening in the second/third/first gap, yielding a nontrivial FNTI II/III/IV with $q = \pm i/\pm j/\pm k$. Further phase transition leads to the FNTI V/VI/VII, which exhibits distinct TEM configurations despite belonging to the same conjugacy class of III/IV/II. The subsequent phase transition culminates in the anomalous FNTI (VIII) that hosts TEMs in all three gaps, despite having a trivial charge of $q = +1$. The dashed arrows highlight a transition path from the trivial to the anomalous FNTI, which experiences band inversions across all three gaps. **b** Multifold bulk-edge correspondence interpretated by the phase-band singularities of the time-evolution operator over one driving period: the system's bulk charge ($q$) equals the ordered product of all charges ($\bar{q}_m$) of phase-band singularities, each of which determines a TEM in the corresponding gap. This is exemplified by the anomalous FNTI, where the three TEMs result from

three phase-band closings. Note that the phase bands at $t = T$ exactly correspond to the system's quasienergy bands.

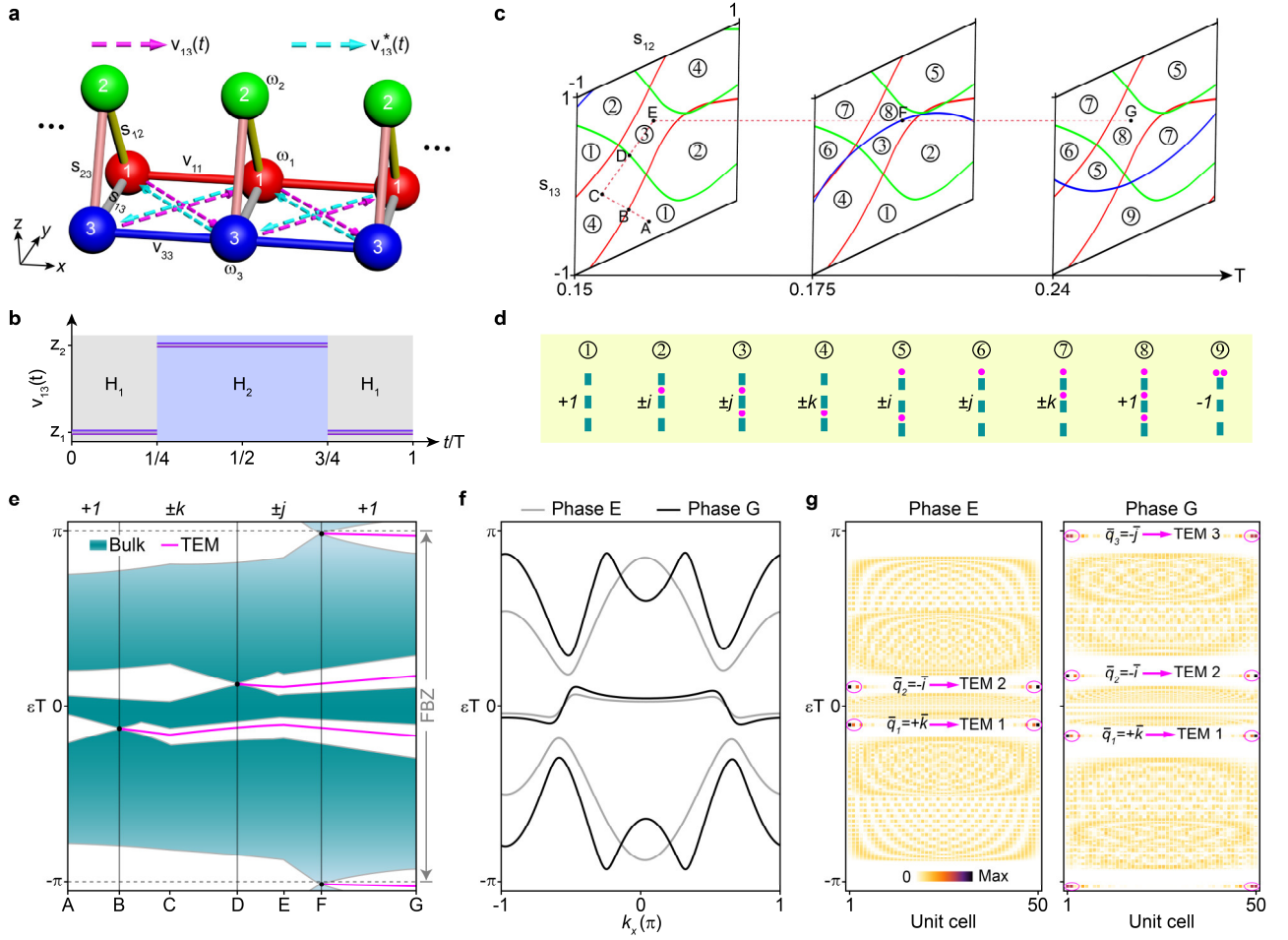

**Fig. 2 | Tight-binding model. a** Schematic of our 1D three-band Floquet model with $\mathcal{PT}$ symmetry. Each unit cell contains three sites (colored spheres) connected by static real couplings (solid rods) and dynamic complex couplings (dashed arrows). **b** $\mathcal{PT}$-symmetric driving protocol. **c** Phase diagram in the parameter space $(s_{12}, s_{13}, T)$, where red, green, and blue solid lines indicate the closures of the first, second, and third bandgaps, respectively. **d** Quaternion charges and TEM configurations of nine representative phases, further concretizing the multifold bulk-edge correspondence. **e** Spectral evolution of a finite-sized lattice along the transition path A → G sketched by the dashed polyline in **c**, which resembles the path highlighted by dashed arrows in Fig. 1a. **f** Quasienergy band structures for phases E and G. **g** Corresponding eigenstate distributions of the finite-sized systems. Notably, the anomalous FNTI (phase G) exhibits TEMs in all bandgaps.

**Theoretical Models**

As depicted in Fig. 2a, we begin with a simple 1D three-band dynamic model with $\mathcal{PT}$ symmetry. Its Floquet-Bloch Hamiltonian in momentum space reads

$$H(k_x, t) = \begin{bmatrix} h_{11}(k_x) & s_{12} & h_{13}(k_x, t) \\ s_{12} & \omega_2 & s_{23} \\ h_{13}^*(k_x, t) & s_{23} & h_{33}(k_x) \end{bmatrix}. \quad (1)$$

More concretely, the static, diagonal matrix elements can be written as $h_{11}(k_x) = \omega_1 + 2v_{11} \cos k_x$ and $h_{33}(k_x) = \omega_3 + 2v_{33} \cos k_x$, while the dynamic, off-diagonal one takes the form of $h_{13}(k_x, t) = s_{13} + v_{13}(t)e^{-ik_x} + v_{13}^*(t)e^{ik_x}$. Here, the complex-valued intercell coupling $v_{13}(t)$ is a periodic function of time $t$. As shown in Fig. 2b, we simply adopt a step-like driving scheme for $v_{13}(t)$: $z_1 = 0.5$ if $t/T \in [0, 1/4] \cup [3/4, 1]$ and $z_2 = 0.5 + 0.5\mathrm{i}$ if $t/T \in [1/4, 3/4]$, with $T$ being the driving period. (Thus, the real part of $v_{13}(t)$ is time invariant, while its imaginary part switches between 0 and 0.5.) The other parameters in $H(k_x, t)$ are real constants: intracell couplings $s_{12} = -0.1$, $s_{23} = 0.5$, and $s_{13} = 0.3$; intercell couplings $v_{11} = 1$ and $v_{33} = -1$; and onsite energies $\omega_1 = 0.5$, $\omega_2 = 0$, and $\omega_3 = -0.5$.

To obtain Floquet quasienergy bands and associated eigenstates of the driven system, we consider the effective Floquet Hamiltonian

$$H_F = \mathrm{i} \log U(T)/T, \quad (2)$$

where the Floquet operator $U(T)$ is given by $U(T) = \mathrm{e}^{-\mathrm{i}H_1 T/4} \mathrm{e}^{-\mathrm{i}H_2 T/2} \mathrm{e}^{-\mathrm{i}H_1 T/4}$, with real step Hamiltonians $H_1 = H(t = 0)$ and $H_2 = H(t = T/2)$. Solving the eigenvalue problem $H_F |u_n\rangle = \varepsilon_n |u_n\rangle$, we obtain the quasienergies within the first Floquet Brillouin zone (FBZ), $\varepsilon_n \in (-\pi/T, \pi/T]$. Notably, given the symmetric driving protocol in Fig. 2b, the effective Floquet Hamiltonian $H_F$ respects $\mathcal{PT}$ symmetry ($H_F = H_F^*$) and enables real-valued eigenstates $|u_n\rangle$. These eigenstates, in turn, allow for the definition of non-Abelian frame charges that characterize the rotation of the eigenstate frame as the momentum $k_x$ varies from $-\pi$ to $\pi$. The frame charges can be either theoretically calculated by generalized Wilson operators or directly visualized from eigenstate frame spheres[1,27]. As shown in the phase diagram (Fig. 2c), our Floquet model supports all types of frame charges in the quaternion group and hosts a rich variety of TEM configurations (Fig. 2d). Figure 2e shows the spectral evolution of a finite-sized lattice along the dashed polyline path A → G indicated in Fig. 2c. Starting from the trivial phase A, the three bandgaps close and reopen sequentially, during which new TEMs emerge in the corresponding gap. Accordingly, the bulk quaternion charge evolves from $+1$ to $\pm k$ and $\pm j$, and finally back to $+1$ (see *Supplementary Fig. 1*). Intriguingly, without any static analog, the final phase G hosts TEMs in all gaps despite a trivial bulk. More concretely, Figs. 2f and 2g showcase comparative quasienergy bands and finite-system eigenstates for phases E and G.

Consistent with the predictions based on phase-band singularities (see *Supplementary Fig. 2*), phase G, the anomalous FNTI, exhibits TEMs in all bandgaps, in contrast to phase E, where the $\pi$-gap TEM disappears.

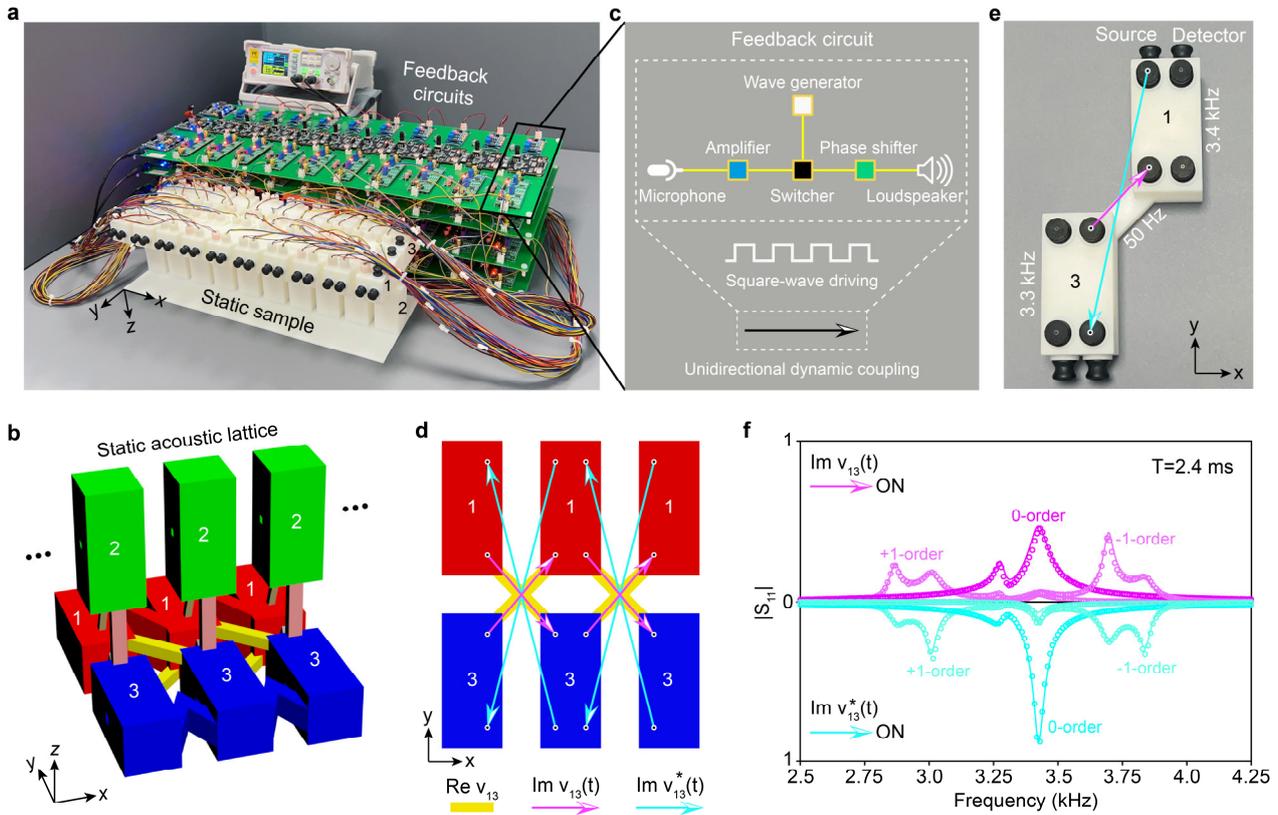

**Fig. 3 | Acoustic emulation of the FNTI model. a** Experimental setup. The static sample, together with feedback circuits for dynamic couplings, realizes our acoustic Floquet lattice. **b** Geometry structure of the static sample. The air-filled cavities emulate atomic orbitals and the narrow tubes mimic static couplings between them. **c** Circuit realization of a unidirectional, square-wave dynamic coupling between cavities 1 and 3, illustrated by the black arrow. **d** Connections of the active circuits used for realizing the dynamic couplings $\mathrm{Im}(v_{13})$ and $\mathrm{Im}(v_{13}^*)$. Note that the yellow tubes contribute the static couplings $\mathrm{Re}(v_{13}) = \mathrm{Re}(v_{13}^*)$. **e** Binary cavity-tube structure used to characterize the unidirectional dynamic couplings. **f** Transmission spectra $|S_{11}|$ (circles) measured by individually activating the coupling circuits of $\mathrm{Im}(v_{13})$ and $\mathrm{Im}(v_{13}^*)$. All spectral responses for the 0 and $\pm 1$ orders are plotted against the excitation frequency. The emergence of high-order harmonic signals (amplified by a factor of 20), as predicted by Floquet theory (lines), witnesses the implementation of dynamic couplings.

*Acoustic implementation of three-band Floquet lattices*

Now, we turn to the acoustic realization of the anomalous FNTI. Figure 3a shows our experimental setup. It consists of a static acoustic lattice (10 unit cells in total) and active feedback circuits that implement unidirectional, time-periodic acoustic couplings[45,46]. As displayed in Fig. 3b, our static sample is realized by an air-filled cavity-tube structure, where the cavities simulate the lattice sites with onsite-energies $\omega_1 = 3400$ Hz, $\omega_2 = 3350$ Hz, and $\omega_3 = 3300$ Hz (effectively, $\omega_1 = 50$ Hz and $\omega_3 = -50$ Hz if shifting $\omega_2$ to 0). The narrow tubes mimic the couplings between cavities: $s_{12} = -10$ Hz, $s_{23} = 50$ Hz, $s_{13} = 30$ Hz, $v_{11} = 100$ Hz, $v_{33} = -100$ Hz, and the time-independent $\text{Re}(v_{13}) = 50$ Hz. The performance of this static lattice has been examined in our acoustic experiments (see *Supplementary Fig. 3*).

The unidirectional, step-like dynamic couplings $\text{Im}(v_{13})$ and $\text{Im}(v_{13}^*)$ are achieved by the external circuits that connect cavities 1 and 3, as sketched in Figs. 3c and 3d. Experimentally, to realize a unidirectional coupling with desired amplitude and phase, the sound signal in cavity 1 (3) is picked up by a microphone, modulated by an amplifier and a phase shifter, and then fed back to cavity 3 (1) through a loudspeaker[45]. Furthermore, a wave generator provides a $T$-period square-wave voltage to the switcher, ultimately yielding a unidirectional coupling swept between 0 and 50i Hz (or $-50i$ Hz) with a duty cycle of 50% (Fig. 3c). Together with the time-invariant static coupling $\text{Re}(v_{13}) = 50$ Hz, ultimately, we realize the dynamic complex couplings $v_{13}(t)$ and $v_{13}^*(t)$. To demonstrate our acoustic implementation of dynamic coupling, we connect two sets of opposite coupling circuits to a simple binary cavity-tube structure (Fig. 3e). Activating one set of circuits while deactivating the other, we measure the transmission response $|S_{11}|$ by exciting cavity 1 and detecting the pressure signal in the same cavity. Note that under the time-periodic modulation, each single-frequency excitation generates multiple harmonics spaced by $1/T$. As an example (with $T = 2.4$ ms), Fig. 3f shows the 0-order and $\pm 1$-order harmonic components varying with the excitation frequency (rather than with the harmonic frequencies themselves). All experimental data match well with the predictions from temporal coupled-mode theory[47]. It is worth pointing out that the high-order harmonics, originating from the couplings between the high-order Floquet replicas and the 0-order one, are essential for our subsequent experimental characterization of FNTIs.

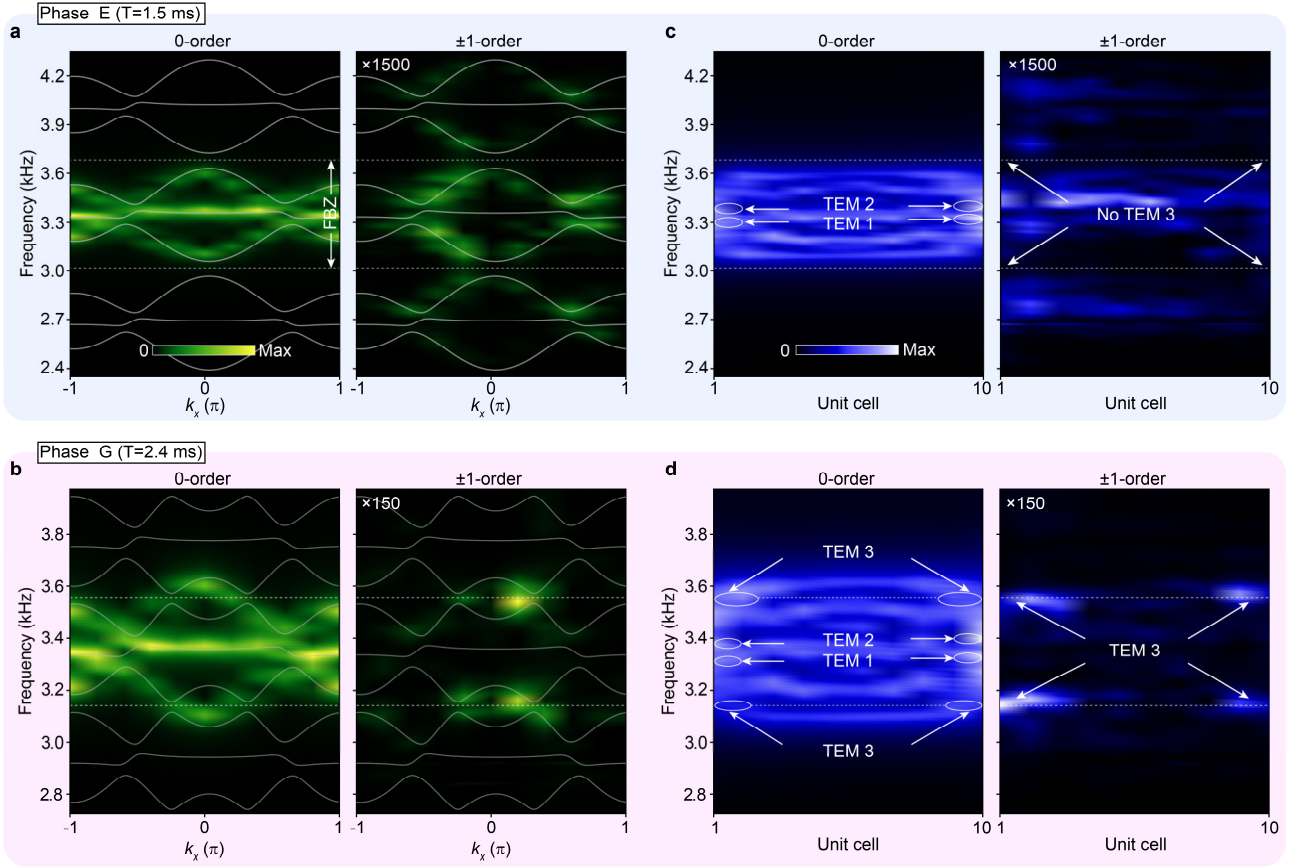

**Fig. 4 | Characterizing Floquet quasienergy bands and TEMs. a,b** Experimentally measured bulk spectra (color scale) for phases E and G, along with their quasienergy band structures (solid lines) for comparison. The horizontal dashed lines indicate FBZ boundaries. To facilitate comparison with the 0-order harmonic signals, the data superimposed from the ±1-order harmonics are amplified by factors of 1500 for phase E and 150 for phase G. **c,d** Frequency-resolved sound distributions detected for phases E and G. In contrast to phase E, phase G exhibits additional $\pi$-gap TEMs in both the 0-order and ±1-order harmonic signals.

## *Bulk and edge responses of anomalous FNTIs*

Next, we characterize the quasienergy bands and TEM patterns for phases E and G, which correspond to the scenarios before and after closing $\pi$-gap. (Note that all acoustic couplings and effective onsite energies are proportional to those used in the tight-binding model.) Conveniently, these two FNTIs can be realized in the same experimental settings, where the phase transition is achieved by simply reconfiguring the driving period $T$ from 1.5 ms to 2.4 ms. To measure the quasienergy spectra, we position a sound source at the middle cell of the sample and scan the sound pressure response across all unit cells. As mentioned above, the presence of high-order harmonic signals serves as a distinctive signature that distinguishes Floquet systems from static systems. To avoid mixing different harmonics, we employ single-frequency excitation and extract the data from the 0-order and $\pm 1$-order harmonics separately. After performing a spatial Fourier transform, we obtain the momentum-space sound energy distribution for each harmonic at given excitation frequency. Finally, sweeping frequency gives the bulk spectra of different harmonics. Furthermore, to characterize the associated TEM configurations, we apply onsite excitation-detection across all cavities in the sample, during which the movable sound source and probe are positioned in the same cavity.

Figures 4a and 4b present our experimentally measured quasienergy spectra (color scale) for phases E and G, respectively. In both cases, the 0-order harmonic spectra match well the predicted band structures (solid lines). Interestingly, compared to phase E, phase G exhibits a significantly stronger high-order harmonic signal around the $\pi$-gap. (Note that a much smaller amplification factor is applied to phase G.) This is because the $\pi$-gap in phase G is caused by band inversion, where neighboring Floquet replicas exhibit strong interactions near the gap. Figures 4c and 4d display the frequency-dependent sound distributions measured for phases E and G. In both cases, the 0-order data show TEMs inside gaps 1 and 2. The key distinction of phase G (anomalous FNTI) from phase E is the emergence of TEM 3 in the $\pi$-gap. This is demonstrated more clearly in the high-order results, where phase G exhibits four strong $\pi$-gap bright spots at both ends of the sample. Overall, our experimental data for quasienergy spectra and TEM patterns align well with coupled-mode theory predictions (see *Supplementary Fig. 4* and *Supplementary Fig. 5*).

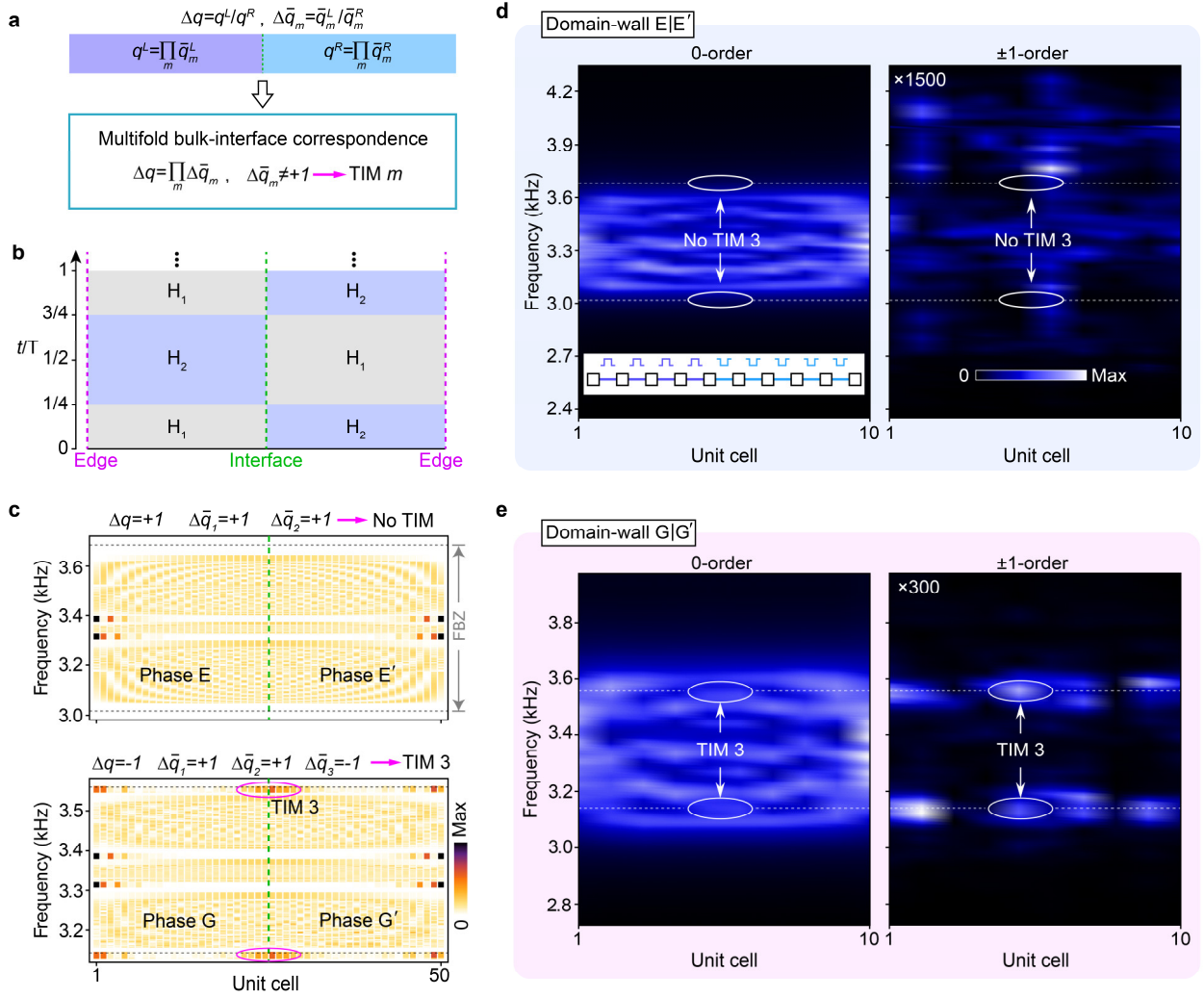

**Fig. 5 | Observation of counter-intuitive TIMs in domain-wall systems. a** Multifold bulk-interface correspondence manifested through the multiplicative relation $\Delta q = \prod_m \Delta \bar{q}_m$, each nontrivial $\Delta \bar{q}_m$ enabling a TIM in gap $m$. **b** Schematic of the domain-wall system formed by two FNTIs with swapped driving sequences. **c** Eigenfield patterns of the domain-wall systems E|E′ and G|G′. The latter exhibits a TIM 3 in the $\pi$-gap due to the nontrivial $\Delta \bar{q}_3 = -1$. **d** Sound distribution characterized for the domain-wall system E|E′. The inset illustrates swapped square-wave sequences in the left and right subsystems. **e** Similar to **d**, but for the system G|G′, which clearly shows the emergence of the $\pi$-gap TIM 3.

## Exotic TIMs induced by swapped driving sequences

Akin to the multifold bulk-edge correspondence, a domain-wall system formed by two distinct FNTIs follows a multifold bulk-interface correspondence. It is governed by the multiplicative relation $\Delta q = \prod_m \Delta \bar{q}_m$, each nontrivial $\Delta \bar{q}_m$ enabling topological interface modes (TIMs) inside gap $m$. As illustrated in Fig. 5a, here $\Delta q = q^L/q^R$ and $\Delta \bar{q}_m = \bar{q}_m^L/\bar{q}_m^R$ respectively characterize charge variations of the quasienergy bands and phase-band singularities between the left and right subsystems. In this work, we are particularly interested in the domain-wall system comprising two FNTIs with swapped driving sequences, which can support exotic TIMs not accessible in Floquet Abelian systems, thereby serving as another unique manifestation of the Floquet non-Abelian topology[12].

As depicted in Fig. 5b, the left and right subsystems follow the driving sequences $H_1 \to H_2 \to H_1$ and $H_2 \to H_1 \to H_2$ within one full period, respectively. Their Floquet operators can be related by a unitary transformation $U^L = V^{-1} U^R V$, with $V = e^{-iH_2 T/4} e^{-iH_1 T/4}$ accounting for a $T/2$ time shift. Despite sharing identical bulk spectra, the different Floquet eigenstates enable either equal or opposite quaternion charge for each phase-band singularity, i.e., $\Delta \bar{q}_m = \pm 1$. More concretely, we construct the domain-wall systems E|E′ and G|G′ based on the existing FNTIs E and G, where the charge variations are outlined in Fig. 5c (see details in *Supplementary Fig. 6*). As expected, in contrast to the system E|E′ that shows no TIM in any gap, the system G|G′ displays the emergence of TIM 3 inside the $\pi$-gap due to the nontrivial charge variation $\Delta \bar{q}_3 = -1$. To experimentally verify these interface phenomena, we implement the domain-wall systems by simply applying a $T/2$ time delay between the square-wave voltage signals of the left and right subsystems. Figures 5d and 5e present the sound distributions measured for these two domain-wall systems. As shown in Fig. 5d, no TIM signals appear in the domain-wall system E|E′, neither in the 0-order nor the high-order data. In contrast, for the system G|G′ in Fig. 5e, the $\pi$-gap TIM 3 signal emerges in the 0-order spectrum and is more clearly discernible in the high-order harmonic data, particularly when compared to the single-crystal case in Fig. 4d. All experimental data closely match the coupled-mode simulations (see *Supplementary Fig. 7*). Essentially, the appearance of the TIM reflects the non-commutativity of the two driving sequences, offering another distinctive manifestation of the Floquet non-Abelian dynamics.

### *Conclusion and outlook*

We have implemented FNTIs in acoustics and unambiguously identified their TEM configurations from both zero-order and high-order harmonic spectra. Notably, the presence of high-order harmonic topological responses serves as a distinctive hallmark of real-time modulated systems. This stands in sharp contrast to the earlier Floquet Abelian systems[35,37,38,41,43,44], where static waveguides are used to emulate time degree of freedom. Moreover, we have demonstrated novel TIMs in a domain-wall system composed of two anomalous FNTIs with swapped driving sequences. Again, this effect is counterintuitive and inaccessible in Floquet Abelian systems[29,30], as the constituent subsystems must share identical Abelian topological invariants[12]. Our real-time experimental setup, with exceptional controllability and reconfigurability, can be adapted to explore a broad range of dynamic topological phenomena, such as Floquet-induced non-Abelian braiding[11], time-modulated non-Hermitian skin effects, and temporal TIMs in momentum gaps[48]. To conclude, this work marks a significant advance in non-equilibrium topological phases and opens the door to future experimental studies of time-varying systems with unprecedented topological dynamics.

*Methods*

Our static acoustic lattice was 3D-printed using photosensitive resin, which provides a rigid boundary condition due to its high impedance. To emulate three desired onsite energies, small square prisms were inserted into cavities 2 and 3 to fine-tune their resonant frequencies, while cavity 1 remained unchanged. Small holes were perforated in each cavity for inserting the sound source and probe, as well as for connecting circuit elements (cavities 1 and 3). To minimize sound leakage, unused holes were sealed with threaded stoppers during the measurements. Furthermore, to achieve time-periodic couplings $\text{Im}(v_{13})$ and $\text{Im}(v_{13}^*)$, 36 sets of carefully calibrated unidirectional coupling circuits are integrated into the static sample, based on the connections outlined in Fig. 3d.

Experimentally, we launched a series of nearly single-frequency sound signals (spaced by 10 Hz) and measured the pressure responses individually. After performing a time-domain Fourier transform, the Floquet modulations produce multiple equidistant harmonic wave signals for each single-frequency incident signal. The collected data were classified according to their harmonic orders and plotted against the excitation frequency (e.g. Fig. 3f). To measure the quasienergy spectra in Figs. 4a,b, a sound source was fixed at the cavity $i$ ($i = 1{\sim}3$) of the middle cell of the sample, along with a probe placed nearby for phase reference. Another probe was moved to scan the sound pressure in cavity $i$ across all unit cells. After normalizing the data to the source signal, a 1D spatial Fourier transform was performed to obtain bulk spectra in momentum space. Finally, the three sets of data were summed in energy to ultimately obtain quasienergy spectra. Similar procedures were applied to characterize TEMs (Figs. 4c,d) and TIMs (Figs. 5d,e), but with onsite excitation and detection across all cavities in the sample.

*Data availability*

The data that support the plots in this paper and other findings of this study are available from the corresponding author upon reasonable request.

*Code availability*

All codes accompanying this publication are directly available from the corresponding author upon reasonable request.

*Acknowledgements*

This project is supported by the National Natural Science Foundation of China (Grants No. 12374418, No. 12404510, and No. 12304495), the National Key R&D Program of China (Grant No. 2023YFA1406900), the Fundamental Research Funds for the Central Universities, the National Postdoctoral Program for Innovative Talents (Grant No. BX20240269), the China Postdoctoral Science Foundation (Grant No. 2024M752454), the Natural Science Foundation of Hubei Province of China (Grant No. 2024AFB064), and the Postdoctoral Project of Hubei Province (Grant No. 2024HBBHCXB056).

*Author contributions*

C.Q. conceived the idea and supervised the project. H.Q. carried out the theoretical analysis and did the simulations. S.T. performed the experiments under the help of Q.Z. and K.Z. H.Q., S.T., and C.Q. analyzed the data and wrote the manuscript. All authors contributed to scientific discussions of the manuscript.

*Competing interests*

The authors declare no competing interests.

**Correspondence** and requests for materials should be addressed to Chunyin Qiu.